\documentclass[pre,preprint]{revtex4}

\usepackage[utf8x]{inputenc}
\usepackage[dvips]{graphicx}
\usepackage{bm}
\usepackage{amssymb}
\usepackage{amsmath}

\newcommand{\bee}{\begin{eqnarray}}
\newcommand{\eee}{\end{eqnarray}}

\begin{document}

\title{Rotational and translational self-diffusion in concentrated suspensions of permeable particles}

\author{Gustavo C. Abade$^1$}

\author{Bogdan Cichocki$^2$}

\author{Maria~L.~Ekiel-Je\.zewska$^3$} \email{mekiel@ippt.gov.pl}

\author{Gerhard N\"agele$^4$}

\author{Eligiusz Wajnryb$^3$}

\affiliation{$^1$ Departamento de Engenharia Mec\^anica, Faculdade de Tecnologia,
Universidade de Bras\'{\i}lia, Campus Universit\'ario Darcy Ribeiro,
70910-900 Bras\'{\i}lia-DF, Brazil}

 \affiliation{$^2$ Institute of Theoretical Physics, Faculty of Physics, University of Warsaw, Ho\.za 69,
  00-681 Warsaw, Poland}

 \affiliation{$^3$ Institute of Fundamental Technological Research,
             Polish Academy of Sciences, Pawi\'nskiego 5B, 02-106 Warsaw, Poland}

 \affiliation{$^4$ Institute of Complex Systems (ICS-3), Research Centre J\"ulich, D-52425 J\"ulich, Germany}

\date{\today}

\begin{abstract}

In our recent work on concentrated suspensions of uniformly porous colloidal spheres with excluded volume interactions, a variety of
short-time dynamic properties were calculated, except for the rotational self-diffusion coefficient.
This missing quantity is included in the present paper.
Using a precise hydrodynamic force multipole simulation method, the rotational self-diffusion coefficient is evaluated for concentrated suspensions of permeable particles. Results are presented for particle volume fractions up to 45\%,
and for a wide range of permeability values. From the simulation results and earlier results for the first-order virial coefficient, we find that
the rotational self-diffusion coefficient of permeable spheres can be scaled to the corresponding
coefficient of impermeable particles of the same size.
We also show that a similar scaling applies to the translational self-diffusion coefficient considered earlier.
From the scaling relations, accurate analytic approximations for the rotational and translational self-diffusion coefficients
in concentrated systems are obtained, useful to the experimental analysis of permeable-particle diffusion.
The simulation results for rotational diffusion of permeable particles are used to show that a generalized Stokes-Einstein-Debye
relation between rotational self-diffusion coefficient and high-frequency viscosity is not satisfied.

\end{abstract}
\maketitle

\section{Introduction}

The rotational and translational self-diffusion of interacting colloidal and macromolecular particles suspended in a
low-molecular-weight solvent is the subject of ongoing research both experimentally and theoretically \cite{PuseyReview:91,NaegeleBook:05}.
Originally, self-diffusion in dilute systems was studied, however the center of interest has shifted since to concentrated dispersions where solvent-mediated
many-particle hydrodynamic interactions (HIs) are of central importance. An example, of biological relevance, is self-diffusion of
proteins and other macromolecules in the crowded environment of a cell \cite{Ando:10}.

Two central quantities quantifying the configuration-averaged influence of HIs on the suspensions dynamics
are the concentration-dependent short-time rotational and translational self-diffusion coefficients $D_r$ and $D_t$, respectively.
At zero particle concentration, these quantities reduce to the single-particle diffusion coefficients $D_0^r$ and $D_0^t$. For
solvent-impermeable colloidal hard spheres with stick hydrodynamic surface boundary conditions, the single-particle coefficients are given by
\begin{eqnarray}
  D_0^{r,\textrm{hs}}  &=& \frac{k_B T}{8 \pi \eta_0 a^3} \\
  D_0^{t,\textrm{hs}} &=& \frac{k_B T}{6 \pi \eta_0 a} \,,
\end{eqnarray}
with $\eta_0$ the Newtonian solvent shear viscosity, Boltzmann constant $k_B$, temperature $T$ and hydrodynamic particle radius $a$.
The influence of the HIs at non-zero concentrations gives rise to values for $D_r$ and $D_t$ smaller than their respective values $D_0^r$ and $D_0^t$ at infinite dilution.
The short-time coefficients describe self-diffusion on the time scale $t \ll a^2/D_0^t$, but with $t$ large enough
that solvent and particle velocity correlations have decayed. On the colloidal short-time scale,
the concentration dependence of $D_r$ and $D_t$ is determined by averaging the HIs with the equilibrium particle distribution.

Self-diffusion coefficients in colloidal suspensions have been determined experimentally by a variety of techniques. The
mean-squared displacement (MSD) of partially solvent-index matched suspensions of colloidal spheres \cite{vanMegen:86, vanMegen:99}
has been measured as a function of time using dynamic light scattering (DLS), with $D_t$ determined from the initial slope of the MSD.

For the vast majority of systems where this specialized
index-matching technique is not applicable, $D_t$ may be inferred, to decent accuracy according to theory and simulation
\cite{SegreBehrend:95,BanchioASD:08,ACENW-JCP-Hq:10},
from a first cumulant analysis of the scattered light electric field autocorrelation function, probed at a scattering wavenumber larger than the peak location
of the static structure factor where the structure factor attains the value one \cite{Pusey:78,HolmqvistPRL:10,Heinen:10}.
Translational long-time self-diffusion coefficients not considered in the present work can be determined using,
e.g., forced Rayleigh scattering \cite{Dozier:85,Palberg:93}, fluorescence recovery after photobleaching (FRAP) \cite{NaegeleBook:05},
and fluorescence correlation spectroscopy \cite{Wilk:04}.

Experimental studies of rotational colloidal self-diffusion are based on techniques which can distinguish different particle orientations.
Methods which have been used for this purpose are depolarized dynamic light scattering (DDLS) on optically anisotropic particles \cite{Degiorgio:95},
and nuclear magnetic resonance \cite{Kannetakis:97}. More recently developed techniques applicable to a larger variety of systems include
time-resolved phosphorescence anisotropy \cite{Koenderink:01,Koenderink:03} and polarized FRAP \cite{Lettinga:03,Koenderinkthesis:03} measurements.
The latter methods have been carried out using fluorophore-labeled colloidal particles. Most of the published experimental results deal with self-diffusion properties of
monodisperse colloidal systems. However, experimental and theoretical work has been also performed on rotational diffusion in colloidal
mixtures, in particular for binary systems where one component (the tracer) is very dilute \cite{Koenderink:01,Koenderink:03,Koenderinkthesis:03,Zhang:02}.
In addition, DDLS measurements of the rotational diffusion of tracer spheres in a polymeric solution have been used to infer
viscoelastic properties from a frequency-dependent generalized Stokes-Einstein-Debye (GSED) relation \cite{ArauzLara:05}.

From a theoretical viewpoint, short-time rotational self-diffusion in monodisperse colloidal systems of non-permeable spheres was studied
using lattice-Boltzmann (LB) \cite{Hagen:99}, Stokesian dynamics \cite{Phillips:88}, and accelerated Stokesian dynamics (ASD) simulations \cite{BanchioASD:08}. While there has been no theoretical work so far on rotational self-diffusion in dense suspensions of porous particles, other transport
properties of porous particles have been studied, including the high-frequency shear viscosity \cite{MoSangani:94,Potanin:95,ElliotRussel:97,Nomennsen:99},
and to first order in concentration the mean sedimentation velocity \cite{ChenCai:99}.

%
%
%

In our earlier work on the short-time dynamics of concentrated suspensions of uniformly porous particles, a broad spectrum
of dynamic properties has been calculated, including the hydrodynamic function \cite{ACENW-JCP-Hq:10,ACENW-PRE-Hq:10} and sedimentation
coefficient \cite{ACENW-JCP-Hq:10}, translational self-diffusion coefficient \cite{ACENW-JCP-Hq:10,ACENW-PRE-Hq:10},
and the high-frequency-limiting shear viscosity $\eta_\infty$ \cite{ACENW-JCP-Visco:10,ACENW-JPCM-GSE:10}. These simulation studies were amended
by the derivation of easy-to-use approximate analytic expressions of good accuracy, notably a generalized Saito formula for the
shear viscosity \cite{ACENW-JCP-Visco:10}, and a spherical annulus model approximation for $\eta_\infty$ \cite{ACENW-JCP-Visco:10},
and to first order in concentration also for $D_t$ and $D_r$ \cite{CENW2011}.
Additionally, precise values for the first-order virial coefficients
of $D_r$ and $D_t$ corresponding to two-body HIs have been obtained \cite{CENW2011}.

In all these studies on permeable particles, the solvent flow inside the spheres is
described by the Debye-B\"uche-Brinkman (DBB) equation \cite{Brinkman:47d,DebyeBueche:48}, and the particles
are assumed to interact directly by excluded volume (i.e., hard-sphere type) forces.
Our simplifying particle model is specified by two parameters only,
namely the particle volume fraction $\phi = (4\pi/3) n a^3$, where $n$ is the number concentration,
and the ratio, $x$, of the particle radius, $a$, to the hydrodynamic penetration depth, $\kappa^{-1}$, inside a permeable sphere. Large (low) values
of $x$ correspond to weakly (strongly) permeable particles.
Typical values for $x$ in permeable-particle systems,
such as core-shell particles, are in the range of $x \sim 30$ or larger \cite{Nommensen:01b}.
While a specific intra-particle structure is ignored in the
model, it is generic in the sense that a more complex internal hydrodynamic structure can be approximately accounted for in terms of a mean permeability.
Porous-particle systems of current interest include dendrimers \cite{Likos:02,Likos:2010,Adamczyk:04}, microgel particles
\cite{Pyett:05,Richtering:08,Coutinho:08},
core-shell particles \cite{Nommensen:01b,Petekidis:04,Zackrisson:05,Masliyah:87},
and star-like polymers of lower functionality \cite{LikosRichter:98}.

The present work complements our earlier analysis of the short-time dynamics in concentrated suspensions of uniformly
permeable spheres by giving simulation results and a theoretical analysis of the short-time rotational diffusion coefficient
not considered so far at non-dilute concentrations. On employing the multipole simulation method of a very high accuracy \cite{CFHWB:94} encoded in the {\sc hydromultipole} program package \cite{CEW:99},
we calculate the short-time rotational self-diffusion coefficient, $D_r(x,\phi)$, as a function of $x$ and $\phi$.
Our results cover the full range of porosities, with the volume fraction extending up to 0.45. In combination with recently obtained tabulated values for
the first-order virial coefficients of $D_r(x,\phi)$ and $D_t(x,\phi)$ \cite{CENW2011}, and precise {\sc hydromultipole} simulation results for
$D_t(x,\phi)$ obtained earlier \cite{ACENW-JCP-Hq:10,ACENW-PRE-Hq:10}, we show that both $D_r(x,\phi)$ and $D_t(x,\phi)$ can be scaled,
in the whole range of permeabilities and volume fractions, to the self-diffusion coefficients $D_r^\textrm{hs}(\phi) = D_r(\infty,\phi)$ and $D_t^\textrm{hs}(\phi) = D_t(\infty,\phi)$ of non-permeable hard spheres with stick boundary conditions
and the same size. From these scaling relations,
accurate analytic expressions for $D_r(x,\phi)$ and $D_t(x,\phi)$ are obtained. We expect these expressions to be useful in the experimental data analysis of diffusion measurement
on permeable particle systems. The present simulation results for $D_r(x,\phi)$, and known results for $\eta_\infty(x,\phi)$, are used
to show the violation of a GSED relation between $D_r(x,\phi)$ and $\eta_\infty(x,\phi)$, amending our earlier study
of similar GSE relations in \cite{ACENW-JPCM-GSE:10}.

The paper is organized as follows:
Sec. \ref{sec:RotationalDiffusion} provides the theoretical background on short-time self-diffusion of permeable particles. Furthermore,
it includes our simulation results for rotational self-diffusion. The scaling relations allowing to map
permeable to non-permeable hard-sphere systems are discussed in Sec. \ref{sec:Scaling}. In Sec. \ref{sec:ZeroPermeability},
we complete the scaling relations by providing simple expressions for the scaling functions for non-permeable hard spheres.
We also discuss the
special case of non-permeable hard spheres in comparison to earlier simulations and experimental work.
In Sec. \ref{sec:GSED}, we demonstrate the violation of the GSED relation.
In our conclusions in Sec. \ref{sec:Conclusions}, we explicitly write convenient expressions
for $D_r(x,\phi)$ and $D_t(x,\phi)$ which should prove useful in practical applications.

\section{Short-time rotational self-diffusion: theory and results}
\label{sec:RotationalDiffusion}

Like in our earlier work on the dynamics of permeable particle systems \cite{ACENW-JCP-Hq:10,ACENW-PRE-Hq:10,ACENW-JCP-Visco:10,ACENW-JPCM-GSE:10,
CENW2011}, we employ a model of uniformly permeable spheres of radius $a$,
dispersed in a Newtonian fluid of viscosity $\eta_0$. The low-Reynolds number incompressible flow inside and outside the spheres
is described, respectively, by the Stokes \cite{HappelBrenner:book,KimKarrila:book}
and Debye-Bueche-Brinkman \cite{Brinkman:47d,DebyeBueche:48} equations
\begin{eqnarray}
\label{eq:DBB}
    \eta_0\;\! {\bm \nabla}^2 {\bm v}({\bm r}) - \eta_0\;\! \kappa^2 \;\!\chi({\bm r})\left[{\bm v({\bm r})} -
    {\bm u}_i({\bm r}) \right]  - {\bm \nabla} p({\bm r})  = 0 \,.
\end{eqnarray}
Here, ${\bf v}$ and ${\bf p}$ are the fluid velocity and pressure, respectively, and $\kappa^{-1}$ is the hydrodynamic
penetration depth. The characteristic function, $\chi({\bf r})$, is equal to one for the field point ${\bf r}$
inside any of the spheres and zero outside. The skeleton of a particle $i$, centered at ${\bf r}_i$, moves rigidly with the local
velocity ${\bf u}_i({\bf r}) = {\bf U}_i + {\bm{\omega}}_i \times \left( {\bf r} - {\bf R}_i \right)$, determined
by the translational and rotational velocities ${\bf U}_i$ and ${\bm{\omega}}_i$, respectively. The
fluid velocity and stress change continuously across a particle surface.

The short-time rotational self-diffusion coefficient of a quiescent, isotropic system is given
in frame-invariant notation by \cite{CEW:99,Zhang:02}
\begin{eqnarray}
\label{eq:Drot}
    D_r \;\!= \;\! \frac{k_B T}{3}\;\! {\Big<} \frac{1}{N} \sum_{i=1}^N \textrm{Tr}\;\! \bm{\mu }_{ii}^{rr}({\bf X}) {\Big>} \,,
\end{eqnarray}
where ${\bf X} = \{{\bf r}_1,\cdots,{\bf r}_N\}$ is the configuration of $N \gg 1$ sphere centers,
and \textrm{Tr} denotes the trace operation.
The hydrodynamic mobility tensor, $\bm{\mu }_{ii}^{rr}({\bf X})$, linearly relates the torque acting on a particle $i$ to its
rotational velocity, for zero forces and torques exerted on the other particles. For the present model system,
the average $\langle \cdots \rangle$ is taken over an equilibrium distribution of non-overlapping spheres,
consistent with the periodic boundary conditions used in our simulations.
Our numerical calculation of $D_r(x,\phi)$ makes use of Eq. (\ref{eq:Drot}).

The coefficient $D_r$ is a function both of $x$ and $\phi$. At infinite dilution, Eq. (\ref{eq:Drot}) reduces to
\cite{FelderhofDeutch:75,Reuland:78}
\begin{equation}
  D_{0}^{r}(x) \;\!=\;\! \frac{k_B T}{8 \pi \eta_0 a^3 \left[  1 + \dfrac{3}{x^2} - \dfrac{3 \coth x}{x} \right] }\,.
\label{eq:Dzerorot}
\end{equation}
Note here that $D_{0}^{r}(x) > D_0^{r,\textrm{hs}}$ unless $x = \infty$.

We have calculated $D_r(x,\phi)$ to high precision
using a hydrodynamic multipole method corrected for lubrication \cite{CichockiFelderhofSchmitz:88,CFHWB:94,CEW:99,CJKW:00},
and encoded in the {\sc hydromultipole} program package extended to permeable spheres.
The hydrodynamic particle structure enters into the {\sc hydromultipole} method only through a single-particle friction operator,
whose form is known for a variety of particle models \cite{FelderhofDeutch:75,CichockiFelderhofSchmitz:88,Reuland:78}.
The details of the simulation method are given elsewhere \cite{ACENW-JCP-Hq:10}.
The values for $D_r$ presented in the following have been determined from equilibrium configuration averages
over typically $N = 256$ particles
in a periodically replicated cubic simulation box, using 100 initial random configurations for each set
of parameters. This gives a statistical
relative error of less than 0.001. In our multipole expansion method used for the rotational mobility tensor
in Eq. (\ref{eq:Drot}), the multipole order, $L$, was truncated usually at $L = 3$.
To gain high-precision data, extrapolations to $L \to 8$ were made, leading to an accuracy in $D_r$ better
than 1\%. The calculated values for $D_r(N)$ using the periodic simulation box with $N$ particles are not critically dependent on the system size,
since $D_r(N=\infty)-D_r(N)$ scales with the particle number like $1/N$. This system size dependence is similar to that of the
high-frequency viscosity, $\eta_\infty(x,\phi)$, of permeable particles. The latter was
calculated in earlier work \cite{ACENW-JCP-Visco:10,MoSangani:94}.
\begin{table}[t]
\caption{Simulation results for the normalized short-time rotational self-diffusion coefficient $D_r(x,\phi)/D^r_0(x)$.}
\label{tab1}
\begin{tabular}{cccccccc}
\hline \hline
$\phi \setminus x$&5&10&20&30&50&100&$\infty$\\
\hline
0.05\;\;&  0.995&  0.987&  0.980&  0.977&  0.973&  0.970&  0.967\\
0.15\;\;& 0.983& 0.958& 0.934& 0.922& 0.911& 0.901& 0.888\\
0.25\;\;& 0.968& 0.925& 0.881& 0.860& 0.839& 0.820& 0.796\\
0.35\;\;& 0.951& 0.886& 0.820& 0.788& 0.757& 0.729& 0.690\\
0.45\;\;& 0.932& 0.842& 0.753& 0.711& 0.669& 0.629& 0.576\\
\hline \hline
\end{tabular}
\end{table}

Table \ref{tab1} lists our high-precision simulation results for $D_r(x,\phi)$, for volume fractions up to $\phi = 0.45$.
Values of the inverse (reduced) penetration depth $x$ are considered from
a very small value $ x \sim 5$, characteristic of highly permeable particles, up to $x = \infty$ characteristic of dry particles
with stick surface boundary conditions.

\section{Scaling self-diffusion of permeable to non-permeable particles}
\label{sec:Scaling}

From analyzing the numerical data for the rotational self-diffusion coefficient in Table~\ref{tab1}, we have found an interesting scaling of permeable to non-permeable spheres of the same size. In addition, we found that a similar scaling is valid for the translational self-diffusion coefficient.
Therefore, results for both quantities will be given in this section. We start from a brief comparison of $D_r(x,\phi)$ to $D_t(x,\phi)$.

\begin{figure*}[th]  
\begin{center}
\includegraphics[width=8.1cm]{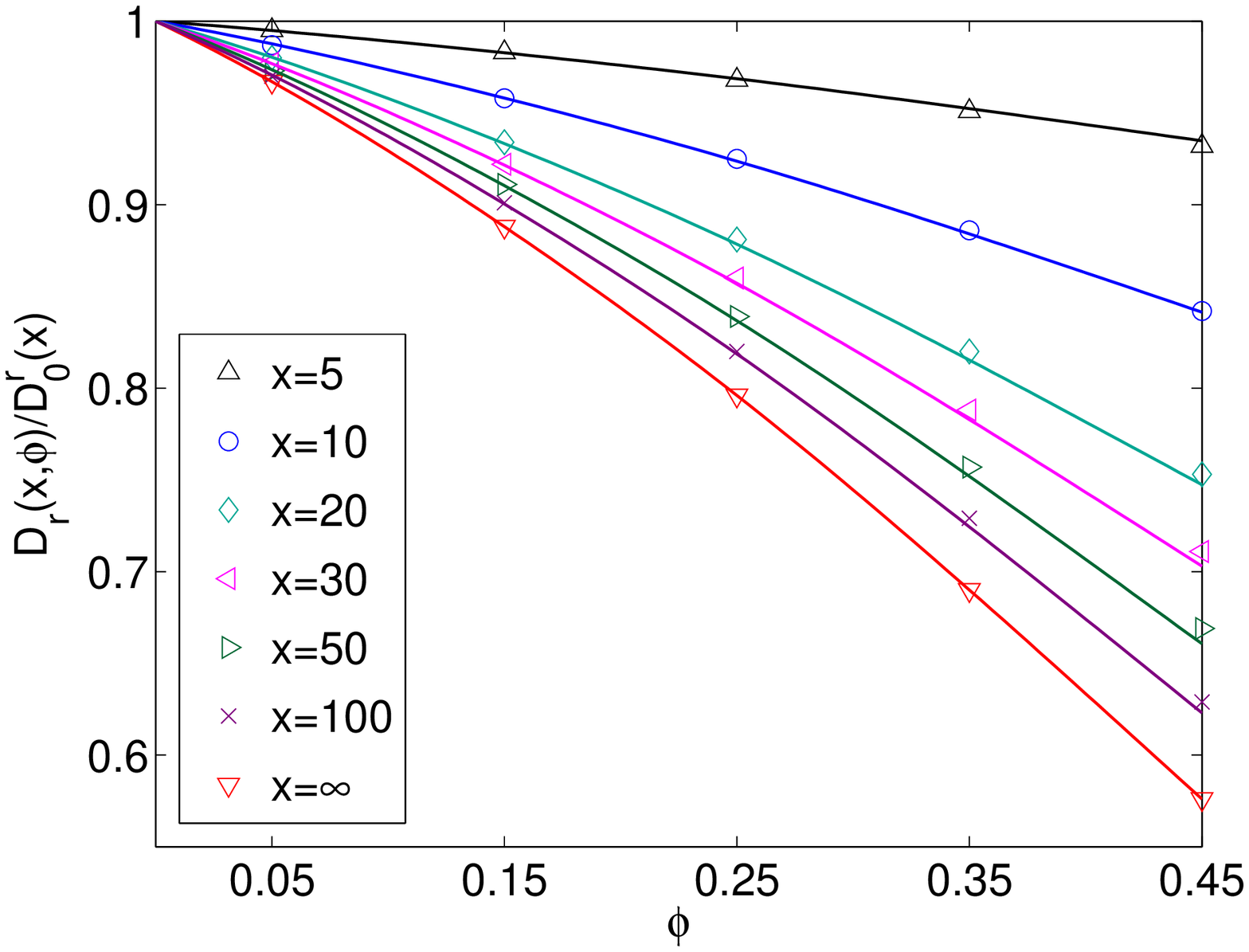}
\includegraphics[width=8.1cm]{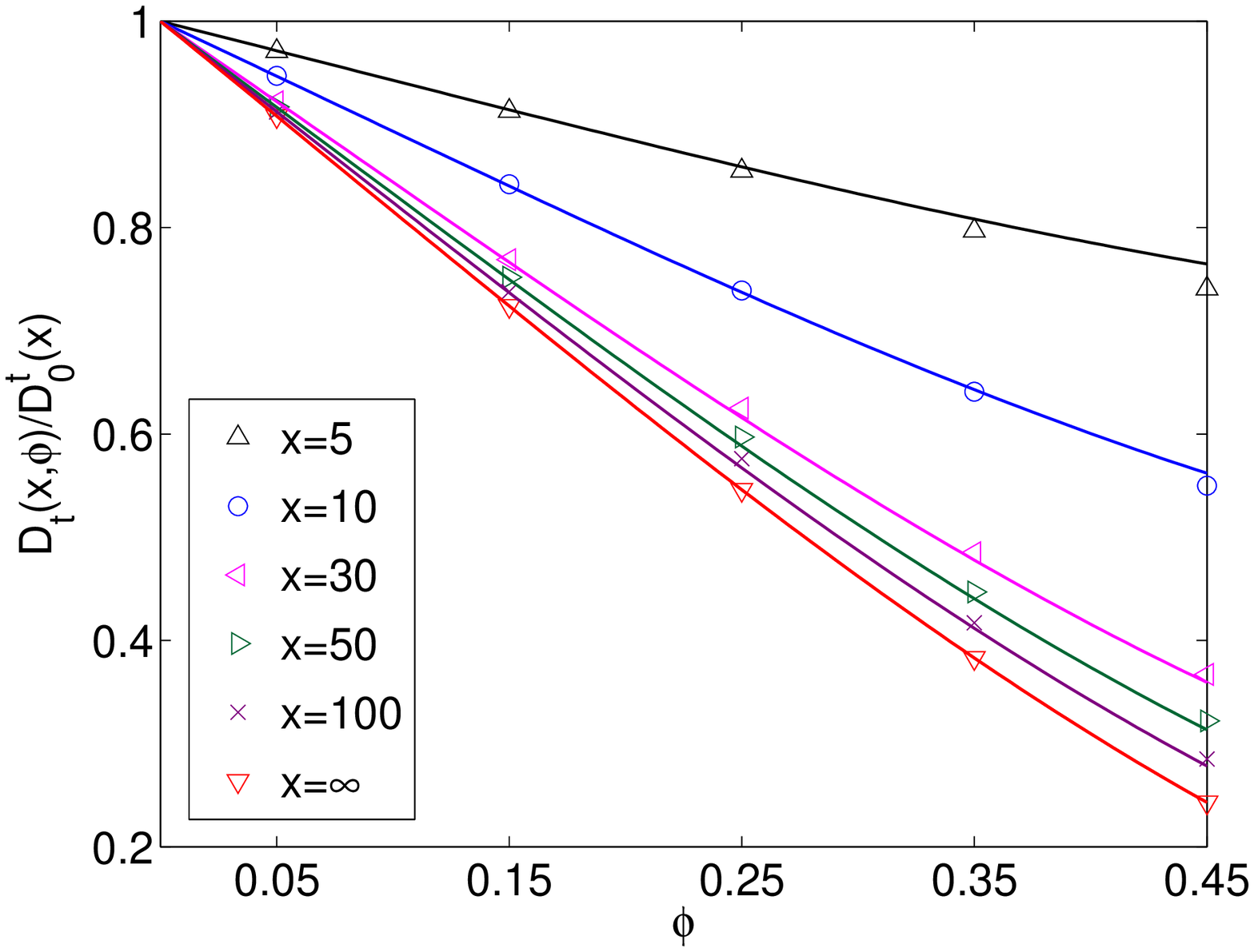}
\caption{Rotational (left) and translational (right) self-diffusion coefficients, $D_r(x,\phi)/D^r_0(x)$ and $D_t(x,\phi)/D^t_0(x)$, as functions of $\phi$, for  values of $x$ as indicated. Symbols: simulation results. Solid lines: interpolated r.h.s. of Eqs.~(\ref{eq:ar}-\ref{eq:at}).
}
\label{fig1}
\end{center}
\end{figure*}

The simulation results for $D_r(x,\phi)$ from table~\ref{tab1} are depicted in the left panel of
Fig.~\ref{fig1} using symbols. For comparison, the right panel of Fig.~\ref{fig1} shows the corresponding simulation results for $D_t(x,\phi)$ taken
from \cite{ACENW-JCP-Hq:10}. For permeable particles, the fluid is allowed to penetrate so that the strength of the HIs is
decreasing with increasing permeability, i.e., decreasing $x$. This is the reason for the larger values of $D_r$ and $D_t$ at larger
permeabilities. Our results show that the effect of HIs on $D_r$ is weaker than on $D_t$, i.e., for a given $x$ and $\phi$, the
reduction of the self-diffusion coefficient relative the infinite dilution value is smaller for rotational diffusion.

%

The numerical results for $D_r(x,\phi)$ and $D_t(x,\phi)$ plotted in Fig.~\ref{fig1} have significantly different slopes
at small volume fractions $\phi$. On the other hand, these slopes are well-reproduced by the first-order virial coefficients, $\lambda_r(x)$ and
$\lambda_t(x)$, defined by the following relations,
\bee
 \frac{D_r(x,\phi)}{D^r_0(x)} &=& 1+\lambda_r(x) \phi + {\cal O}(\phi^2) \\
 \frac{D_t(x,\phi)}{D^r_0(x)} &=& 1+\lambda_t(x) \phi + {\cal O}(\phi^2) \,,
\eee
evaluated in Ref.~\cite{CENW2011} and listed in Table~\ref{tab2}. The single-particle rotational diffusion coefficient, $D^r_0(x)$,
has been already given in Eq.~\eqref{eq:Dzerorot}, and the translational one has the form given in~\cite{Brinkman:47d,DebyeBueche:48},
\bee
  D_0^t(x) &=&  \frac{k_B T}{6\pi \eta_0 a} \left( 1 + \frac{1}{x\;\!\coth{x} - 1} + \frac{3}{2\;\!x^2} \right) \,.
\eee
%

\begin{table}[h]
\caption{First-order virial terms, $\lambda_r(x)$ and $\lambda_t(x)$, of the rotational and translational self-diffusion coefficients
\cite{CENW2011}.}
\label{tab2}
\begin{tabular}{clllllll}
\hline
&\,\,\,\,\,\,5&\,\,\,\,10&\,\,\,\,20&\,\,\,\,30&\,\,\,\,50&\,\,\,\,100&\,\,\,\,\,$\infty$ \\ \hline \hline
$\lambda_r$ & \,\,-0.097 & -0.236 & -0.376 & -0.442 & -0.505 & -0.561 & -0.631 \\
$\lambda_t$ & \,\,-0.569 & -1.060 & -1.416 & -1.550 & -1.661 & -1.746 & -1.832 \\ \hline \hline
\end{tabular}
\end{table}

\begin{figure*}[th]  
\begin{center}
\includegraphics[width=8.1cm]{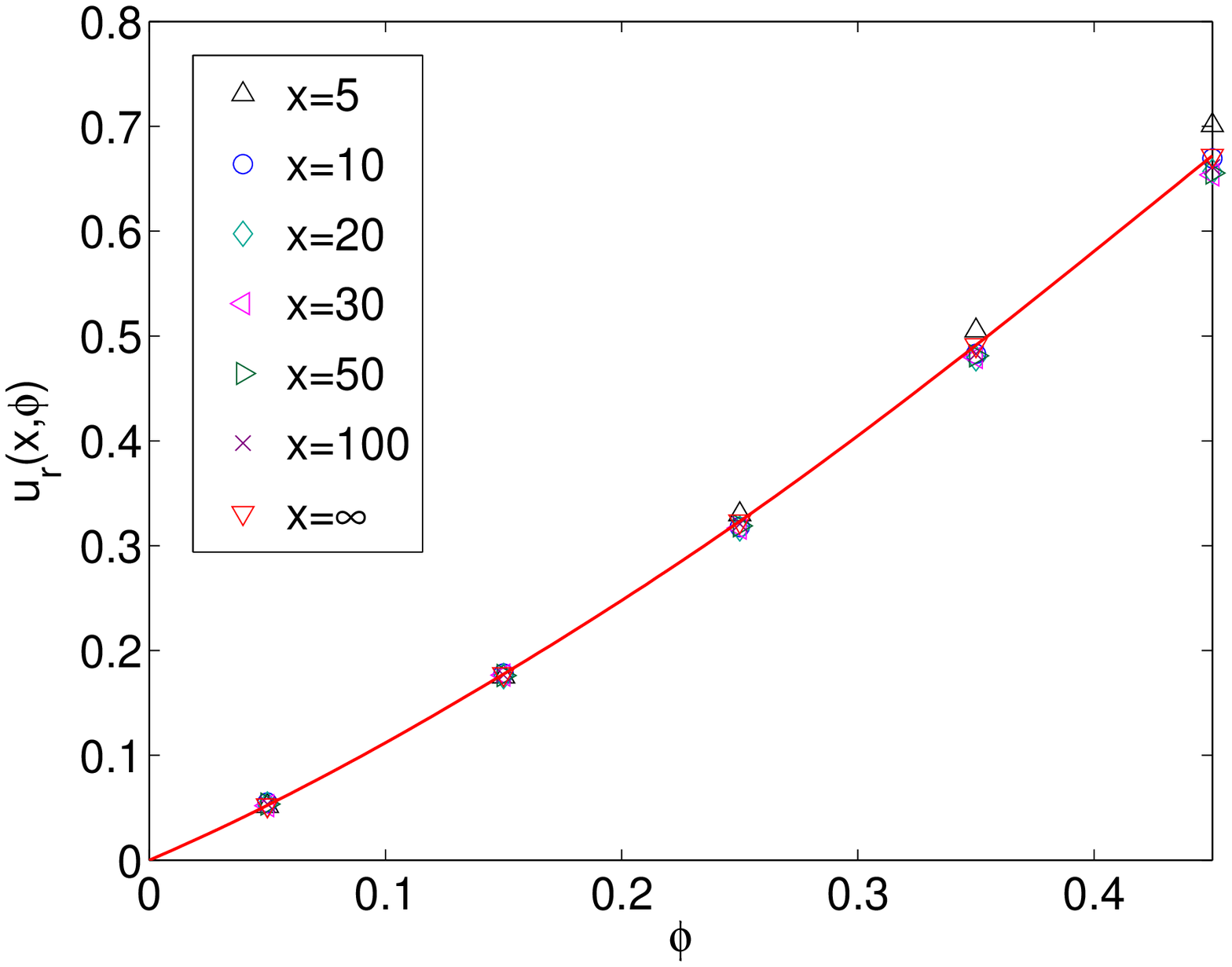}
\includegraphics[width=8.1cm]{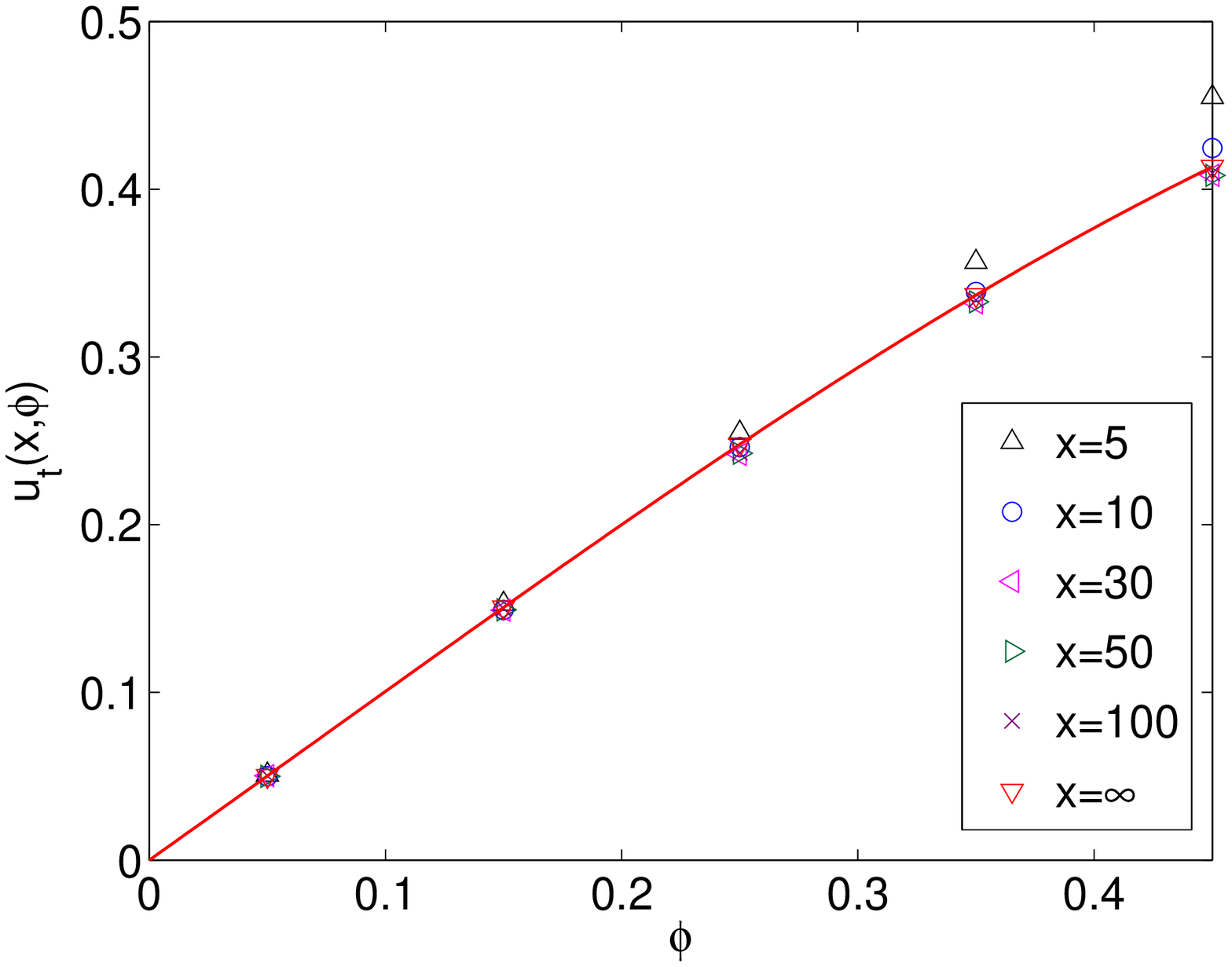}
\caption{The functions $u_r(x,\phi)$  and $u_t(x,\phi)$ are practically independent of $x$.
Symbols: simulation results for the indicated values of $x$. Solid lines: spline fit interpolations of $u_r(\infty,\phi)$ and
$u_t(\infty,x)$.
}
\label{fig2}
\end{center}
\end{figure*}

Therefore, the idea is to introduce the following scaling functions,
\bee
  u_r(x,\phi) &=&\left(\frac{D_r(x,\phi)}{D^r_0(x)}- 1 \right) \frac{1}{\lambda_r(x)} \,,
\label{ueq}\\
  u_t(x,\phi) &=&\left(\frac{D_t(x,\phi)}{D^t_0(x)}- 1 \right) \frac{1}{\lambda_t(x)} \,.
\label{ueqt}
\eee
For all values of $x\ge 5$, the functions $u_r(x,\phi)$ and $u_t(x,\phi)$ do practically not depend on $x$, i.e.
they are permeability-independent. Indeed,
as shown in Fig. \ref{fig2}, the curves for $u_r$ and $u_t$  as functions of $\phi$
collapse on the corresponding curves for the non-permeable solid spheres, i.e.
\bee
  u_r(x,\phi) &\approx& u_r(\infty,\phi)\,,  \label{uuu} \\
  u_t(x,\phi) &\approx& u_r(\infty,\phi)\,,\label{uuut}
\eee
%
%
with a relative error less than 3\% for  $x\ge 10$.

%

Therefore, the short-time self-diffusion coefficients in suspensions of
permeable particles are well approximated by the following expressions,
%
\bee
  \frac{D_r(x,\phi)}{D^r_0(x)} &\approx& 1 +\lambda_r(x)u_r(\infty,\phi) \,, \label{eq:ar}\\
  \frac{D_t(x,\phi)}{D^t_0(x)} &\approx& 1 +\lambda_t(x)u_t(\infty,\phi)  \,.     \label{eq:at}
\label{dyrr}
\eee
In Fig. \ref{fig1}, the solid, continuous lines are not just mere fits to the simulation data, but represent
the expressions in Eqs. (\ref{eq:ar}-\ref{eq:at}), i.e. the outcome
of the interesting scaling behavior of the short-time self-diffusion of permeable particles.
The error made in using Eqs. (\ref{eq:ar} -\ref{eq:at}) instead of the precise simulation values, is at most 1\% for
rotational and 3\% for translational self-diffusion. To complete the analysis, we need to specify the scaling functions for the non-permeable solid spheres. This will be done in the next section.

\section{Self-diffusion coefficients of non-permeable spheres}
\label{sec:ZeroPermeability}

We will now use the existing data for non-permeable hard spheres to construct simple approximate expressions for the
scaling functions $u_r(\infty,\phi)$ and $u_t(\infty,\phi)$.

We start with a comparison between our present simulation
results for $D_r^\textrm{hs}(\phi) = D_r(\infty,\phi)$ and $D_t^\textrm{hs}(\phi) = D_t(\infty,\phi)$ for vanishing permeability,
and a selection out of a large body of
published experimental (see, e.g.,\cite{vanMegen:86,Degiorgio:95,Kannetakis:97,Koenderink:03,SegreBehrend:95})
and theoretical (see, e.g., \cite{Hagen:99,Phillips:88,BanchioASD:08,MoSangani:94,Ladd:90,SierouBrady:01,BanchioBrady:03,BradyOpinion:96})
data on impermeable hard spheres.

Related to this comparison, we note first that Cichocki et al. have derived precise second-order virial expansion results \cite{CEW:99}
\begin{eqnarray}
  \frac{D_r^\textrm{hs}}{D_0^{r,\textrm{hs}}} &=& 1 - 0.631 \phi - 0.726 \phi^2  + {\cal O}(\phi^3) \,, \label{eq:vihr}\\
  \frac{D_t^\textrm{hs}}{D_0^{t,\textrm{hs}}} &=& 1 - 1.8315 \phi - 0.219 \phi^2 + {\cal O}(\phi^3)\,,  \label{eq:viht}
\end{eqnarray}
for the short-time rotational and translational self-diffusion coefficients of neutral hard spheres as functions of $\phi$.

\begin{figure*}[ht]
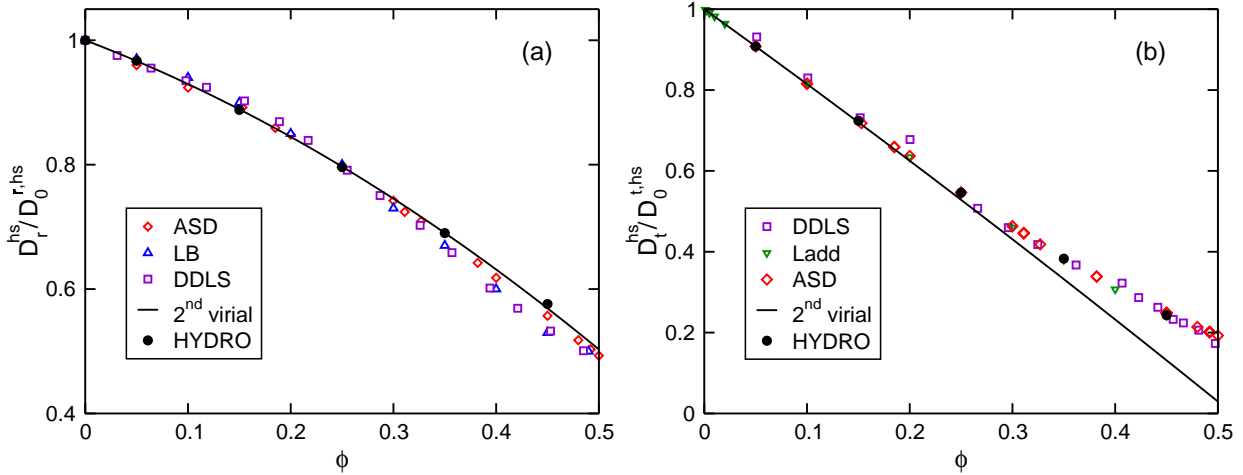
  
\begin{center}
\includegraphics[width=8.1cm]{Abade_fig3a.ps}
\includegraphics[width=8.1cm]{Abade_fig3b.ps}
\caption{(a) Rotational and (b) translational self-diffusion coefficients
of non-permeable hard spheres with stick boundary conditions, as functions of $\phi$. Compared in (a) are our {\sc hydromultipole} data
(labeled HYDRO)
for $D_r^\textrm{hs}/D_0^\textrm{hs}$ with SD \cite{BanchioASD:08} and LB \cite{Hagen:99} simulation results, and DDLS
experimental data \cite{Degiorgio:95} for optically anisotropic particles.
In (b), we compare the {\sc hydromultipole} data for $D_t^\textrm{hs}(\phi)/D_0^{t,\textrm{hs}}$ with ASD simulation results \cite{BanchioASD:08}, force
multipole calculations by Ladd \cite{Ladd:90}, and DLS experimental
data by Segre et al. \cite{SegreBehrend:95}. Solid lines: 2nd-order virial expansion results, in (a) according to Eq. (\ref{eq:vihr}), and in (b) according to Eq. (\ref{eq:viht}).
}
\label{fig3}
\end{center}
\end{figure*}

Regarding rotational diffusion, Fig. \ref{fig3}(a) shows the comparison of our data with Lattice-Boltzmann
\cite{Hagen:99} and ASD \cite{BanchioASD:08} simulation results, and DDLS experimental data \cite{Degiorgio:95}
for optically anisotropic fluorinated polymer particles. The rotational diffusion coefficient as a function of $\phi$ has
a concave shape, different from that for $D_t$ which is weakly convex.
Our simulation data for non-permeable particles agree well with the ASD result.
The LB data at large $\phi$ are somewhat smaller.
The key message conveyed by Fig. \ref{fig3}(a) is that
the second-order virial result for $D_r^\textrm{hs}(\phi)$ in Eq. (\ref{eq:vihr}) describes the simulation and experimental data
remarkably well for all volume fractions up to the freezing transition value $0.49$ \cite{BanchioASD:08,Hagen:99},
indicating that higher-order virial coefficients are small or mutually cancel out.
Therefore, for constructing a simple approximation for $u_r(\infty,\phi)$ from Eq. (\ref{ueq}), it is sufficient to take
as $D_r(\infty,x)$ the 2nd-order virial expansion in Eq. (\ref{eq:vihr}).
In this way, the rotational scaling function is approximated by
\bee
  u_r(\infty,\phi) &\approx& \phi+ 1.151 \;\! \phi^2 \,,
\label{eq:uu}
\eee
with an accuracy of 1.5\% or better relative to our simulation data.

In Fig. \ref{fig3}(b), we compare our simulation data for $D_t^\textrm{hs}(\phi)/D_0^{t,\textrm{hs}}$
with ASD simulation \cite{BanchioASD:08} and force
multipole calculation results \cite{Ladd:90}, and with DLS experimental
data \cite{SegreBehrend:95}. The figure shows that
the translational second-order virial expression in Eq. (\ref{eq:viht}) for $D_t^\textrm{hs}$ noticeably underestimates
the simulation and experimental data when $\phi$ is larger than 0.3.

For this reason, we need a more precise expression for $D_t(\infty,\phi)$ than the 2nd-order virial expansion
in Eq. (\ref{eq:viht}). We have found that our simulation data are approximated with a
0.4\% accuracy by the following expression for the scaling function $u_t(\infty,\phi)$, defined in Eq.~\eqref{ueqt},
\bee
  u_t(\infty,\phi) &\approx& \phi+ 0.12\;\!\phi^2 - 0.65\;\!\phi^3 \,.
\label{eq:uut}
\eee
The term $\phi + 0.12\phi^2$ follows from the virial expansion in Eq. \eqref{eq:viht}, and the coefficient of the 
third order term, $- 0.65\;\!\phi^3$, has been obtained by fitting to the numerical data for $u_t(\infty,\phi)-\phi - 0.12\phi^2$, in the range $0\le \phi\le0.45$.

\section{Generalized Stokes-Einstein-Debye relation}
\label{sec:GSED}

We proceed with the discussion of a generalized short-time GSED relation.
Having obtained in this paper precise numerical data for $D_r(x,\phi)$, and taking values of $\eta_\infty(x,\phi)$
tabulated in \cite{ACENW-JCP-Visco:10},
we are in the position to test the validity of the following short-time GSED relation
\begin{equation}
\label{eq:GSED}
  \frac{D_r(x,\phi)}{D_0^r(x)}\;\!\frac{\eta_\infty(x,\phi)}{\eta_0} \;\! \stackrel{?}{\approx} \;\!1 \,,
\end{equation}
between the rotational self-diffusion coefficient, and the high-frequency viscosity
of permeable particles. The validity of generalized Stokes-Einstein relations such as the present one is an
important issue in microrheological studies where one tries to infer rheological properties more easily from diffusion measurements.
The GSED relation in Eq. (\ref{eq:GSED}) was shown before to be violated for suspensions of non-permeable neutral and charged
particles \cite{Koenderink:03,Banchio:99}. Here, we ask the same validity question for permeable particle systems.
\begin{figure*}[b]  
\begin{center}
\includegraphics[width=8.1cm]{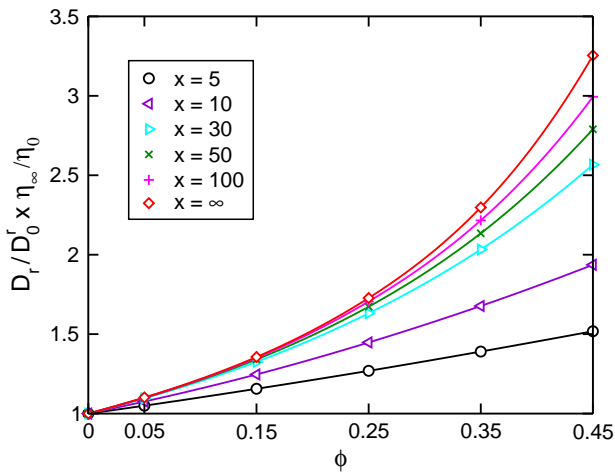}
\caption{The generalized Stokes-Einstein-Debye relation between $D_r(x,\phi)$ and high-frequency viscosity $\eta_\infty(x,\phi)$
is not satisfied. Solid lines are interpolating spline fits to our simulation results (symbols).
}
\label{fig4}
\end{center}
\end{figure*}

In Fig. \ref{fig4}, the GSED relation is examined for different values of $x$. If valid, all
curves should collapse on a single horizontal line of unit height. One notices from the figure that the GSED relation is significantly
violated for $x \geq 30$, and volume fractions $\phi > 0.15$ where the particles
are significantly correlated. Thus, a rotating particle experiences its neighborhood not just as a structureless
medium characterized by the viscosity $\eta_\infty(x,\phi)$.
The GSED relation for rotational diffusion is more strongly violated than its translational
counterpart. As shown in \cite{ACENW-JPCM-GSE:10}, $(D_t/D_0^t)\times(\eta_\infty/\eta_0)$ increases
practically linearly in $\phi$, even for non-permeable particles, whereas in Fig. \ref{fig4} a pronounced non-linear increase is observed.

\section{Conclusions}
\label{sec:Conclusions}

Using the {\sc hydromultipole} simulation method, the short-time rotational self-diffusion coefficient, $D_r(x,\phi)$,
of uniformly permeable spheres was calculated to high precision as a function of permeability and volume fraction.

An interesting scaling relation was found between $D_r(x,\phi)$ and the corresponding coefficient, $D_r(\infty,\phi)$,
of non-permeable, solid spheres of the same size, where the permeability enters only
through the first-order rotational virial coefficient.
A similar scaling was found for translational self-diffusion.

The combination of the scaling relations with accurate 2nd-order and 3rd-order concentration expansion results in Eqs. (\ref{eq:uu}-\ref{eq:uut})
for $D_r(\infty,\phi)$ and $D_t(\infty,\phi)$, respectively, has led us to the expressions
\begin{eqnarray}
  \frac{D_r(x,\phi)}{D^r_0(x)} & \approx & 1 + \lambda_r(x) \;\! \phi \;\! \left( 1 + 1.151 \;\! \phi \right) \,, \\
  \frac{D_t(x,\phi)}{D^t_0(x)} & \approx & 1 + \lambda_t(x) \;\! \phi \;\! \left( 1 + 0.12  \;\! \phi - 0.65 \;\! \phi^2 \right) \,,
\end{eqnarray}
for the permeability-dependent self-diffusion coefficients.
In combination with table \ref{tab2} for $\lambda_r$ and $\lambda_t$,
these are convenient expressions useful in diffusion measurement analysis of permeable particle systems, and
as input to theories of long-time dynamic properties.
The accuracy of these expressions is better than 1.5\% for rotational and 3.5\% for translational self-diffusion,
for the whole range of volume fractions $\phi \le 0.45$ provided $x\ge 5$.
We expect the
expressions to be useful in the experimental analysis of self-diffusion, to gain a quick estimate
of the mean porosity in concentrated systems. Moreover, they can serve as short-time inputs
into theoretical methods of calculating frequency-dependent and long-time diffusion properties, such as in mode-coupling
and dynamic density functional theory methods.

The simulation results for $D_r(x,\phi)$, and recent
results for $\eta_\infty(x,\phi)$, were used to scrutinize the validity of a generalized Stokes-Einstein-Debye
relation in its dependence on permeability. We found this relation to be significantly violated for non-dilute
suspensions, unless the permeability is unrealistically large. The GSED test
for porous particles presented in this paper complements earlier GSE performance tests \cite{ACENW-JPCM-GSE:10}
of different short-time diffusion properties. Of all considered GSE relations,
only the one for the cage diffusion coefficient can claim a certain validity when applied to neutral
porous particles \cite{ACENW-JPCM-GSE:10}. However, also this
relation becomes invalid when the particles are significantly charged \cite{BanchioASD:08}.

With the present paper on self-diffusion in combination with earlier simulation results for other dynamic properties
such as the hydrodynamic function and viscosity, and the development of accurate analytic approximations
for these properties \cite{ACENW-JCP-Hq:10,ACENW-JCP-Visco:10,ACENW-JPCM-GSE:10,CENW2011},
we have obtained an essentially complete description of the short-time dynamics of uniformly
permeable particles with no-overlap interactions.

Dispersions of spherical particles with more complex internal hydrodynamic structure,
such as core-shell particles, and different direct interactions, will be the subject of a future study. \\ \\
\vspace{0.1cm}

\begin{acknowledgments}
M.L.E.J. and E.W. were supported in part by the Polish Ministry of Science and Higher Education grant N N501 156538.
G.N. thanks M. Heinen for helpful discussions and the Deutsche
Forschungsgemeinschaft (SFB-TR6, project B2) for financial support.
Numerical simulations were done at NACAD-COPPE/UFRJ in Rio de 
Janeiro, Brazil, and at the Academic Computer Center in Gdansk, Poland.  
\end{acknowledgments}



\begin{thebibliography}{99}

\bibitem{PuseyReview:91} P.N. Pusey, in {\em Liquids, Freezing, and the Glass Transition},
edited by J. P. Hansen, D. Levesque, and J. Zinn-Justin, Elsevier, Amsterdam, (1991).

\bibitem{NaegeleBook:05} G. N\"agele, J.K.G. Dhont, and G. Meier in {\em Diffusion in Condensed Matter},
edited by P. Heitjans and J. K\"arger, Springer, Berlin (2005).

\bibitem{Ando:10} T. Ando and J. Skolnick, PNAS Early Edition, www.pnas.org/cgi/doi/10.1073/pnas.1011354107 (2010).

\bibitem{vanMegen:86} W. van Megen, S.M. Underwood, and I. Snook, J. Chem. Phys. {\bf 85}, 4065 (1999).

\bibitem{vanMegen:99} A. Brand, H. Versmold, and W. van Megen, J. Chem. Phys. {\bf 110}, 1283 (1999).

\bibitem{SegreBehrend:95} P. N. Segr\`e, O. P. Behrend and P. N. Pusey, {Phys. Rev. E} {\bf 52}, 5070 (1995).

\bibitem{BanchioASD:08} A.J. Banchio and G. N\"agele, J. Chem. Phys. {\bf 128}, 104903 (2008).
\bibitem{ACENW-JCP-Hq:10} G.C. Abade, B. Cichocki, M.L. Ekiel-Je\.zewska, G. N\"agele, and E. Wajnryb, J. Chem. Phys. {\bf 132}, 014503 (2010).

\bibitem{Pusey:78}
P.N. Pusey, {J. Phys. A: Math. Gen.} \textbf{11}, 119 (1978).

\bibitem{HolmqvistPRL:10} P. Holmqvist and G. N\"agele, Phys. Rev. Lett. {\bf 104}, 058301 (2010).

\bibitem{Heinen:10} M. Heinen, P. Holmqvist, A.J. Banchio and G. N\"agele, J. Appl. Cryst. {\bf43}, 970 (2010).

\bibitem{Dozier:85} W.D. Dozier and P.M. Chaikin, J. Phys. (Paris) {\bf46}-C.3, 165 (1985).

\bibitem{Palberg:93} R. Simon, T. Palberg, and P. Leiderer, J. Chem. Phys. {\bf 99}, 3030 (1993).

\bibitem{Wilk:04} A. Wilk, J. Gapinski, A. Patkowski, and R. Pecora, J. Chem. Phys. {\bf 121}, 10794 (2004).

\bibitem{Degiorgio:95} V. Degiorgio, R. Piazza, and R.B. Jones, Phys. Rev. E {\bf 52}, 2707 (1995).

\bibitem{Kannetakis:97} J. Kannetakis, A. T\"olle, and H. Sillescu, Phys. Rev. E {\bf 55}, 3006 (1997).

\bibitem{Koenderink:01} G.H. Koenderink, H. Zhang, M.P. Lettinga, G. N\"agele and A.P. Philipse, Phys. Rev. E {\bf 64}, 2001 (2001).

\bibitem{Koenderink:03} G.H. Koenderink, H. Zhang, D.G.A.L. Aarts, M.P. Lettinga, A.P. Philipse,
and G. N\"agele, Faraday Discuss. {\bf 123}, 335 (2003).

\bibitem{Lettinga:03} M.P. Lettinga, G.H. Koenderink, B.W.M. Kuipers, E.Bessels, and A.P. Philipse, J. Chem. Phys. {\bf 120}, 4517 (2003).

\bibitem{Koenderinkthesis:03} G.H. Koenderink, Ph.D. thesis: {\em Rotational and translational diffusion in colloidal
mixtures}, Utrecht University, The Netherlands (2003).

\bibitem{Zhang:02} H. Zhang and G. N\"agele, J. Chem. Phys. {\bf 117}, 5908 (2002).

\bibitem{ArauzLara:05} E. Andablo-Reyes, P. Diaz-Leyva, and J.L. Arauz-Lara, Phys. Rev. Lett. {\bf 94}, 106001 (2005).

\bibitem{Hagen:99} M.H.J. Hagen, D. Frenkel and C.P. Lowe, Physica A {\bf 272}, 376 (1999).

\bibitem{Phillips:88} R.J. Phillips, J.F. Brady, and G. Bossis, Phys. Fluids {\bf 31}, 3462 (1988).

\bibitem{MoSangani:94} G. Mo and A. S. Sangani,
{Phys. Fluids} {\bf 6}, 1637 (1994).

\bibitem{Potanin:95} A.A. Potanin and W.B. Russel, {Phys. Rev. E} {\bf 52}, 730 (1995).

\bibitem{ElliotRussel:97} S.L. Elliot and W.B. Russel,
{J. Rheol.} {\bf 42}, 361 (1998).

\bibitem{Nomennsen:99}
P. A. Nommensen, M.H.G. Duits, D. van den Ende, and J. Mellema, Phys. Rev. E {\bf 59}, 3147 (1999).

\bibitem{ChenCai:99} S.B. Chen and A. Cai, {J. Colloid Interface Sci.} {\bf 217},
328 (1999).

\bibitem{ACENW-PRE-Hq:10} G.C. Abade, B. Cichocki, M.L. Ekiel-Je\.zewska, G. N\"agele, and E. Wajnryb, Phys. Rev. E. {\bf 81}, 020404(R) (2010).

\bibitem{ACENW-JPCM-GSE:10} G.C. Abade, B. Cichocki, M.L. Ekiel-Je\.zewska, G. N\"agele, and E. Wajnryb, J. Phys.: Condens. Matter {\bf 22}, 322101 (2010).

\bibitem{ACENW-JCP-Visco:10} G.C. Abade, B. Cichocki, M.L. Ekiel-Je\.zewska, G. N\"agele, and E. Wajnryb, J. Chem. Phys. {\bf 133}, 084906 (2010).

\bibitem{CENW2011} B. Cichocki, M.L. Ekiel-Je\.zewska, G. N\"agele and E. Wajnryb, submitted to Phys. Fluids (2011).

\bibitem{Brinkman:47d} H.C. Brinkman, Proc. R. Dutch Acad. Sci. {\bf 50}, 618, 821 (1947).

\bibitem{DebyeBueche:48} P. Debye and A.M. Bueche, J. Chem. Phys. {\bf 16}, 573 (1948).

\bibitem{Nommensen:01b} M.H.G. Duits, P. A. Nommensen, D. van den Ende, and J. Mellema, Colloids Surf. A {\bf 183-185}, 335 (2001).


\bibitem{Likos:02} C. N. Likos, S. Rosenfeldt, N. Dingenouts, M. Ballauff, P. Lindner, N. Werner, and F. V\"ogtle,
J. Chem. Phys. {\bf 117}, 1869 (2002).

\bibitem{Likos:2010} S. Hui{\ss}mann, A. Wynveen, C.N. Likos, and R. Blaak, J. Phys.: Condens. Matter {\bf 22}, 232101 (2010).

\bibitem{Adamczyk:04} Z. Adamczyk, B. Jachimska, and M. Kolasinska, J. Colloid Interface Sci. {\bf 273}, 668 (2004).

\bibitem{Pyett:05} S. Pyett and W. Richtering, J. Chem. Phys. {\bf 122}, 034709 (2005).
\bibitem{Richtering:08} T. Eckert and W. Richtering, {J. Chem. Phys.} {\bf 129}, 124902 (2008).

\bibitem{Coutinho:08} C.A. Coutinho, R.K. Harrinauth, and V.K. Gupta, Colloids Surf. A {\bf 318}, 111 (2008).

\bibitem{Petekidis:04} G. Petekidis, J. Gapi\'nski, P. Seymour, J. S. van Duijneveldt, D. Vlassopoulos
and G. Fytas, {Phys. Rev. E} {\bf 69}, 042401 (2004).

\bibitem{Zackrisson:05} M. Zackrisson, A. Stradner, P. Schurtenberger and J. Bergenholtz, {Langmuir} {\bf 21}, 10835 (2005).

\bibitem{Masliyah:87} J. Masliyah, G. Neale, K. Malysa, and T.G.M. van de Ven, Chem. Eng. Science {\bf 42}, 245 (1987).

\bibitem{LikosRichter:98} C. N. Likos, H. L\"owen, M. Watzlawek, B. Abbas, O. Jucknischke, J. Allgaier, and D. Richter,
Phys. Rev. Lett. {\bf 80}, 4450 (1998).


\bibitem{CFHWB:94}
B. Cichocki, B.U. Felderhof, K. Hinsen, E. Wajnryb, and J. B\l awzdziewicz, J. Chem. Phys. {\bf 100}, 3780 (1994).

\bibitem{CEW:99} B. Cichocki, M.L. Ekiel-Je\.zewska and E. Wajnryb,  {J. Chem. Phys.} {\bf 111}, 3265 (1999).

\bibitem{HappelBrenner:book}
J. Happel and H. Brenner, \textit{Low Reynolds Number Hydrodynamics}, Martinus Nijhoff, Dordrecht (1986).

\bibitem{KimKarrila:book}
S. Kim and S.J. Karrila, \textit{Microhydrodynamics: Principles and Selected Applications}, Butterworth-Heinemann, London (1991).

\bibitem{FelderhofDeutch:75} B.U. Felderhof and J.M. Deutch, {J. Chem. Phys.} {\bf 62}, 2391 (1975).

\bibitem{Reuland:78} R. Reuland, B.U. Felderhof and R.B. Jones, {Physica A} {\bf 93}, 465 (1978).

\bibitem{CichockiFelderhofSchmitz:88} B. Cichocki, B. U. Felderhof and R. Schmitz,
{PhysicoChemical Hydrodynamics} {\bf 10}, 383 (1988).

\bibitem{CJKW:00} B. Cichocki, R.B. Jones, R. Kutteh and E. Wajnryb,
{J. Chem. Phys.} {\bf 112}, 2548 (2000).


\bibitem{Ladd:90} A.J.C. Ladd, {J. Chem. Phys.} {\bf 93}, 3484 (1990).

\bibitem{SierouBrady:01} A. Sierou and J.F. Brady, {J. Fluid
Mech.} {\bf 448}, 115 (2001).

\bibitem{BanchioBrady:03} A.J. Banchio and J.F. Brady, {J. Chem. Phys.} {\bf 118}, 10323
(2003).

\bibitem{BradyOpinion:96} J.F. Brady, {Curr. Opin. Colloid
Interface Sci.} {\bf 1}, 472 (1996).

\bibitem{Banchio:99} A.J. Banchio, G. N\"agele and J. Bergenholtz, J. Chem. Phys. {\bf 111}, 8721 (1999).

\end{thebibliography}
\end{document}